\documentstyle[osa,aps,prl,epsfig]{revtex}
\begin{document}
\twocolumn[\hsize\textwidth\columnwidth\hsize\csname
@twocolumnfalse\endcsname
\title{
Temperature dependence of Mott transition in VO$_2$ and
programmable critical temperature sensor}
\author{Bong-Jun Kim$^{\ast}$, Yong Wook Lee, Byung-Gyu Chae, Sun Jin Yun, Soo-Young Oh, and Hyun-Tak
Kim}
\address{IT Convergence Components Lab., ETRI, Daejeon 305-350, Republic of
Korea}
\author{Yong-Sik Lim}
\address{Department of Applied Physics, Konkuk University, Chungju,
Chungbuk 380-701, Republic of Korea}

\maketitle{}
\begin{abstract}
The temperature dependence of the Mott metal-insulator transition
(MIT) is studied with a VO$_2$-based two-terminal device. When a
constant voltage is applied to the device, an abrupt current jump
is observed with temperature. With increasing applied voltages,
the transition temperature of the MIT current jump decreases. We
find a monoclinic and electronically correlated metal (MCM) phase
between the abrupt current jump and the structural phase
transition (SPT). After the transition from insulator to metal, a
linear increase in current (or conductivity) is shown with
temperature until the current becomes a constant maximum value
above $T_{SPT}{\approx}$68$^{\circ}$C. The SPT is confirmed by
micro-Raman spectroscopy measurements. Optical microscopy analysis
reveals the absence of the local current path in micro scale in
the VO$_2$ device. The current uniformly flows throughout the
surface of the VO$_2$ film when the MIT occurs. This device can be
used as a programmable critical temperature sensor.~~~~~ PACS
numbers:
71.27. +a, 71.30.+h\\
\\
\end{abstract}
]
\newpage

The first-order Mott discontinuous metal-insulator transition
(MIT) has been studied as a function of temperature in numerous
materials such as Ti$_2$O$_3$, V$_2$O$_3$, and VO$_2$ etc
\cite{Mott}. Almost all have a transition temperature, $T_{MIT}$,
below room temperature except VO$_2$ which has
$T_{MIT}{\approx}$68$^{\circ}$C. In particular, VO$_2$ thin films
were used for fabrication of two- and three-terminal devices
controlled by an electric field \cite{Kim1}. A high-speed Mott
switching device using an abrupt current jump as observed in I-V
measurements was predicted for manufacturing in the nano-level
transistor regime \cite{Kim2,Chudnovski}.

Moreover, Raman experiments\cite{Kim3} for a VO$_2$ film have
showed monoclinic-insulator peaks after the film had undergone an
electric-field-induced transition from an insulator to a metal.
Furthermore, tetragonal-metal peaks have been associated with the
structural phase transition (SPT) above 68$^{\circ}$C. Also no
evidence of phonon softening near the transition temperature has
been found by the temperature dependence of Raman spectra measured
with a VO$_2$ single crystal and a thin film \cite{Petrov}. These
results support the electron correlation model of the MIT.
However, some reports argue that the electric field-induced MIT is
due to Joule heating by current and is accompanied by SPT, and
that, furthermore, the local current path or current filament
formed by the dielectric breakdown\cite{Yamanouchi} can also cause
the jump (MIT). The dielectric breakdown was described by
depinning and the collective transport of charge carriers above a
threshold voltage. Here, we try to elucidate this ambiguity
through the analysis of our present research.

Another interesting aspect in VO$_2$ is that the $T_{MIT}$ can be
modified by doping \cite{Futaki,Rakotoniaina}  and stress
\cite{Muraoka}. VO$_2$ thin films deposited on (001) and (110)
TiO$_2$ substrates showed a modified $T_{MIT}$ of 27 and
96$^{\circ}$C, respectively, where the c-axis length was stressed
by a lattice mismatch between the film and the substrate
\cite{Muraoka}. The modification of the $T_{MIT}$ by doping and
stress is restricted to within a fixed temperature, whereas the
$T_{MIT}$ induced by an electric field linearly depends on the
electric field intensity.

\begin{figure}
\vspace{-1.0cm}
\centerline{\epsfysize=10.0cm\epsfxsize=9.0cm\epsfbox{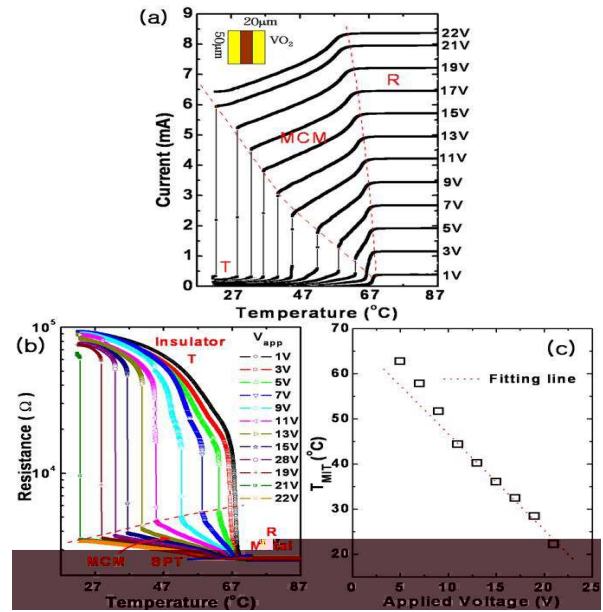}}
\vspace{-0.5cm} \caption{Temperature dependences of, (a) the
current, and (b) the resistance for the MIT (Jump) at applied
voltages. The transition temperatures of the MIT shift towards
room temperature with increasing applied voltages. An intermediate
monoclinic and correlated metal (MCM) phase is located between
red-doted lines. Fig. (b) is plotted from R=V/I in I-T curves of
Fig. 1(a). Fig. (c) The applied voltage dependence of $T_{MIT}$
from fig. (a). Dotted line is a fit to a linear function. The
slope of a fitting line is -2.13$^{\circ}$C/Volt.}
\end{figure}

In this letter, for applications of the MIT over wide temperature
ranges, we observe a modification of the $T_{MIT}$ in a VO$_2$
thin film device by applying voltages. Actually, the modification
of the $T_{MIT}$ is appreciable for a critical temperature sensor
working at any set temperature below 68$^{\circ}$C. The relation
between the MIT and the SPT is investigated by micro-Raman
spectroscopy. By analysis of the micro-scale current localization
with optical microscopy, we confirm a uniform current flow that is
not explained by dielectric breakdown. Moreover, we perform a
switching experiment with a triangle wave to observe generation of
Joule heat in a device.

VO$_2$ films on (10$\bar{1}$0) Al$_2$O$_3$ substrates have been
prepared by the sol-gel method described in detail elsewhere
\cite{Chae1}. In order to observe the temperature dependence of
the MIT, VO$_2$-based two-terminal devices were fabricated in
which VO$_2$ films were ion-milled with an Ar$^{+}$ beam. The
thickness of the VO$_2$ films is about 100 nm. Nickel was used as
an electrode metal for the Ohmic contacts. The fabricated devices
have a channel width of W=50 $\mu$m and a channel length between
electrodes of L=20 $\mu$m. The temperature dependence of the
current was measured from room temperature to 450 K in a cryostat.
The I-V characteristics of the devices were measured by a
precision semiconductor parameter analyzer (HP4156B). As observed
in the I-V measurements, the MIT occurs around 22 V at room
temperature. Thus constant voltages are applied in the range from
1 to 22 V.

Raman spectra were measured with an Ar laser beam to which the
VO$_2$ film between the electrodes was exposed. The 514.5 nm line
of an Ar$^{+}$ laser at a power of 4.5 mW was employed in a
micro-Raman (LABRAM 300) with a spectral resolution less than 2
cm$^{-1}$. The Raman system was also equipped with an Olympus
microscope (BX41) which allowed accurate alignment of the beam
onto the device. During the Raman measurement, the current applied
to the device after the abrupt MIT was limited to the compliance
(restricted) current to prevent any possible damage in the device
due to excess current. The Raman spectrum  measurement took 60
sec.

Figure 1(a) shows the current-voltage curves measured by a VO$_2$
device with $V_{MIT}\approx$21V (inset figure) as a function of
temperature for various constant applied voltages. At
$V_{applied}$ = 1 V, the resistance behavior with the temperature
is typical of a curve with the MIT near 68$^{\circ}$C (Fig. 1b).
Above 68$^{\circ}$C, currents have a constant value of 0.38 mA. As
the constant applied voltage is increased from 1 to 22 V, an
abrupt current jump clearly appear above 5 V. The transition
temperatures gradually shift from 68$^{\circ}$C at $V_{applied}$ =
1 V to room temperature at $V_{applied}$ = 21 V. After an abrupt
current jump occurred in the electric field, VO$_2$ devices showed
the Ohmic behaviors in our previous research \cite{Kim1}.
Moreover, we clearly found a new region of linear behavior of
current in the I-T curves. The region corresponds to an
intermediate regime between abrupt current jumps and a red-dash
line which is the SPT line confirmed by micro-Raman measurements
shown in the next section. Thus we divided the I-T curves into
three phases, the semiconductor-monoclinic transient triclinic T
phase, the intermediate phase and the tetragonal rutile R metal
phase. The SPT temperatures decrease slightly with increasing
applied voltage due to increase of Joule heat.

\begin{figure}
\vspace{-0.8cm}
\centerline{\epsfysize=10cm\epsfxsize=8.5cm\epsfbox{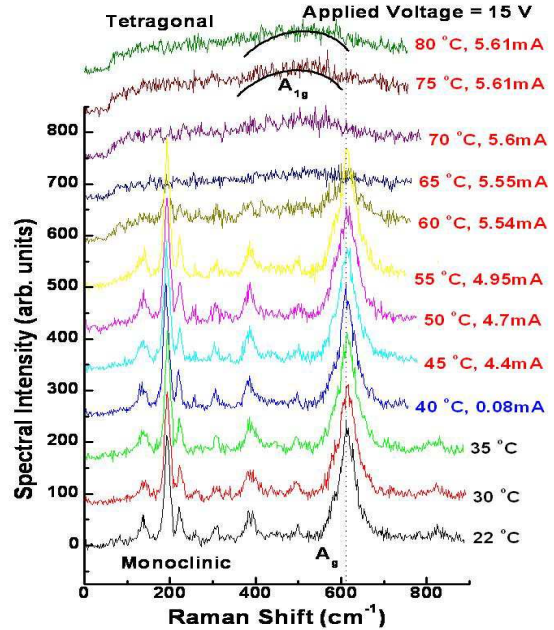}}
\vspace{-0.3cm} \caption{Temperature dependence of Raman peaks at
15 V. The monoclinic A$_g$ peak at 617 cm$^{-1}$ disappears around
65$^{\circ}$C. Above 70$^{\circ}$C, the tetragonal A$_{1g}$ peak
appears and is overlaid with solid lines.}
\end{figure}

Figure 1(b) shows the resistances as a function of temperature at
constant applied voltages and are plotted on a log scale.
Resistances were obtained from R=V/I in Fig. 1(a). In the
intermediate phase between the dashed lines in fig. 1(a), the
current linearly increases with increasing temperature, whereas in
this region the resistance gradually decreases (the conductivity
increases) at each applied voltages, as shown in Fig. 1(b). This
strongly suggests that VO$_2$ is in a metallic state different
from the tetragonal metallic state because the temperature is
still less than 68$^{\circ}$C. Near the SPT, the resistance
becomes a minimum and the conductivity ($\sigma\propto$1/R) has a
maximum value.

In a strongly correlated system, the electrical conductivity is
proportional to square of the effective mass between
quasiparticles; $\sigma\propto(m^{\ast}/m)^2$   where
$m^{\ast}/m=1/(1-(U/U_c)^2)$ for an inhomogeneous system and $U$
is the on-site Coulomb correlation energy between quasiparticles
\cite{Brinkman,Kim4,Kim5}. Therefore, since the intermediate phase
has the maximum conductivity near $T_{SPT}$, it is regarded as the
intermediate phase being strongly correlated; the intermediate
phase is named as a monoclinic and correlated metal (MCM) phase.
The MCM phase arises from inhomogeneity \cite{Kim4,Kim5}.

The MCM phase is described by the equation $n_c(T,E) =  n(E)+
n(T)$, where $n(E)$ is the hole density excited by an electric
field (voltage), $n(T)$ is the hole density excited by
temperature, and $n_c(T,E)$ is the critical hole density in which
the MIT occurs due to the electric field and temperature
excitations \cite{Kim1}. Hole carriers were confirmed by Hall
measurement \cite{Kim1}. For constant $n_c$, $n(T)$ decreases, and
$T_{MIT}$ decreases as $n(E)$ increases, which suggests that the
MIT is controlled by doped holes.

Figure 1(c) shows the applied voltage dependence of $T_{MIT}$
fitting to a linear function; $T_{MIT}$ = -2.13$V_{applied}$ + 68.
This fitting line is denoted as a dotted line. At $V_{applied}$ =
0, $T_{MIT}$ = 68$^{\circ}$C, which is the SPT temperature. Except
for a deviation of the fitting near 68$^{\circ}$C, $T_{MIT}$
follows the linear. This indicates that the device can be used as
a programmable critical temperature sensor.

Figure 2 shows temperature dependence of Raman peaks and currents
simultaneously measured by applying 15 V to the device. The Raman
peaks are measured at a low laser power which was decreased by
using neutral density (ND) filter with an optical density of D=2,
which does not make a local laser beam spot visible on the film
(see next section). In the temperature range from 22 to
40$^{\circ}$C, the intense monoclinic A$_g$ peaks near 193 and 617
cm$^{-1}$ appear in accordance with our previous Raman
experiments. At 45$^{\circ}$C, no clear changes occur in the Raman
curve, nevertheless, the MIT occurred between 40 and 45$^{\circ}$C
with a current jump from 0.08 to 4.4 mA. This indicates that the
MIT occurs without the SPT. Monoclinic A$_g$ peaks are still
visible up to 60$^{\circ}$C with little decrease in intensity.
A$_g$ peaks disappear around 65$^{\circ}$C and the broad
tetragonal A$_{1g}$ peak then appears near 550 cm$^{-1}$. This
indicates the SPT from monoclinic to rutile tetragonal structure.
Above 65$^{\circ}$C, the current has a constant value of 5.61 mA,
as a constant current is shown in the R region of Fig. 1(a). Thus
the dotted red line between MCM and R in Fig. 1(a) is regarded as
the SPT. Note that a difference between
$T_{MIT}\approx$37$^{\circ}$C at 15 V in Fig. 1(a) and
$T_{MIT}\approx$45$^{\circ}$C in Fig. 2 may be due to different
measurement environments in I-V and Raman spectroscopy.

We investigated changes of surface color and surface status of
VO$_2$ by the applied voltages and a laser beam with the variable
temperature. Figure 3(a) shows an optical image of the VO$_2$
device taken at 10 V and 60$^{\circ}$C, in which ND filter was not
used. The dark trace of the laser spot in a dashed circle appears
on VO$_2$ film. The temperature of the film in the laser spot is
higher than 68$^{\circ}$C showing the difference of the
reflectance with another part in the film. At 10 V and
70$^{\circ}$C, laser beam spot is no longer visible, as shown in
Fig. 3(b). The contrast of VO$_2$ film is changed only by the
temperature. Thus the optical image shows that the current flows
uniformly on the VO$_2$ film by applied voltage.

\begin{figure}
\vspace{-0.8cm}
\centerline{\epsfysize=10cm\epsfxsize=8.5cm\epsfbox{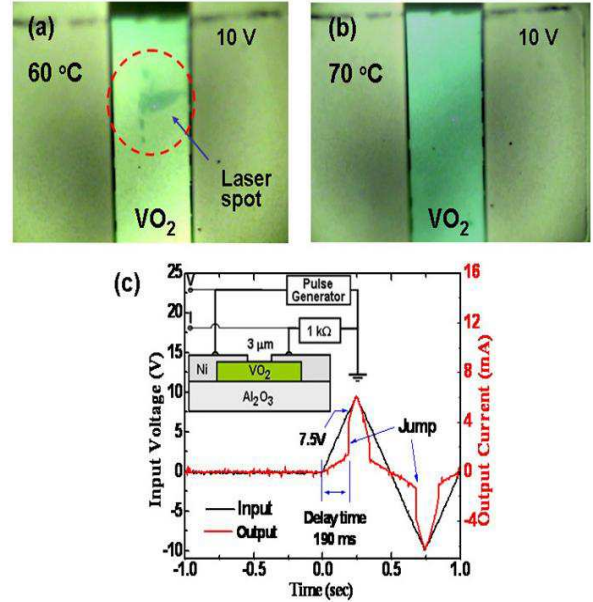}}
\vspace{-0.6cm} \caption{(a) Enlarged images of a VO$_2$ device
taken by optical microscopy. At a high laser power without a ND
filter, a local laser spot appears at 60$^{\circ}$C and 10 V. (b)
At 70$^{\circ}$C and 10 V, the laser spot is no longer visible.
The VO$_2$ films are then in the metal phase. (c) For a VO$_2$
device with L=3 $\mu$m, W=10 $\mu$m. The MIT is switched
repeatedly at 7.5 V (red solid line). The output current (red
right axis) is measured for an input triangle voltage (black solid
line) of 1 Hz (left axis). Delay time $\tau$ took 190 msec.
$V_{MIT}$=7.5V and $I_{MIT}$=4mA were obtained. Inset shows the
measurement circuit.}
\end{figure}

From optical measurements, the transition time of the MIT in
VO$_2$ has been measured to be in the subpicosecond regime
\cite{Cavalleri,Kim6} and in the order of nanosecond for an
electronic device \cite{Chae2}. The heating model predicts that a
delay time takes about 1 $\mu$sec for a device with L=3 $\mu$m and
W=50 $\mu$m to become $T_d=T_{SPT}$ where $T_d$ is a device
temperature in a previous research \cite{Chae2}. In order to check
whether the SPT is produced by Joule heating or not ($Q=
\int^{\tau}_{0}$ IV$dt$, where $\tau\approx$190 msec which is a
long delay time to 1 $\mu$sec; Q increases with increase of
$\tau$.), one triangle wave with a period of 1 sec is applied to a
device with a width of L=3 ${\mu}$m and a length of W=10 ${\mu}$m,
which has $V_{MIT}$ = 7.5 V (Fig. 3(c)). The MIT is shown as a
jump at $V_{MIT}\approx$7.5 V and $I_{MIT}\approx$4 mA. This
indicates that $T_d$ produced by Joule heating is less than
$T_{SPT}$. When $T_d > T_{SPT}$, the current jump should not be
observed because the MIT as observed in the I-V curves is
continuous without jump above $T_{SPT}$ (Fig. 1(b)). Thus Joule
heat does not increase $T_d$ up to $T_{SPT}$, which indicates that
Joule heat is not a cause of the MIT.

In conclusion, the VO$_2$-based devices show the separation of the
MIT from the SPT. The $T_{MIT}$ is controlled by the applied
voltage, a filament for a conducting path is not formed and the
large Joule heat which causes the SPT is not produced even in the
high current induced by the MIT. In future, this device can be
utilized as a programmable critical temperature or infrared sensor
and was named MoBRiK \cite{Kim5}.


\end{document}